\documentclass[aps,nofootinbib,prd,eqsecnum,showpacs,showkeys,preprintnumbers]{revtex4-2}
\usepackage[caption=false]{subfig}
\usepackage{graphicx}
\usepackage{amsmath}
\usepackage{amsfonts}
\usepackage{amssymb}
\usepackage{color}
\usepackage{bm}
\usepackage{mathrsfs}
\usepackage{epstopdf}
\usepackage{url}
\usepackage{footnote}
\usepackage{textcomp}
\usepackage{ulem}
\usepackage{esint}
\usepackage[unicode=true, pdfusetitle,
bookmarks=true,bookmarksnumbered=false,bookmarksopen=false,
 breaklinks=false,pdfborder={0 0 1},backref=false,colorlinks=false]{hyperref}
\usepackage{multirow}
\usepackage{pifont}

\begin{document}
\title{Diagnostic Approaches for Interacting generalized holographic Ricci Dark Energy Models}
	\author{O. Enkhili}
	\email{omarenkhili@gmail.com}
             \author{F. Bargach}
	\email{farida.bargach@gmail.com}
	\author{D. Mhamdi}
	\email{dalale.mhamdi@ump.ac.ma}
\author{T. Ouali}
	\email{t.ouali@ump.ac.ma}
             \author{A. Errahmani}
	\email{ahmederrahmani1@yahoo.fr}
	\date{\today }
	\affiliation{Laboratory of Physics of Matter and Radiations, Mohammed I University, BP 717, Oujda, Morocco}
	\begin{abstract}
                 In this paper, we present an analytical solution for the interacting generalized holographic dark energy model, assuming a linear interaction rate between dark energy and dark matter. We  determine the equation of state parameter, the generalized holographic Ricci dark energy density, the matter density, and the deceleration parameter. By analyzing the behavior of these cosmological parameters, we demonstrate that  our   model   aligns with recent observations and reproduces the late-time accelerated expansion of the Universe. To compare our model with the $\Lambda$CDM model, we use various diagnostic tools including statefinder, $Om(z)$-diagnostic, statefinder hierarchy, growth rate analysis, and $\omega_H$-$\omega'_H$ plane. We also analyze  the stability of the model by examining  the speed of sound.  These methods show  that the dynamics of the Universe remain very close to that of the standard cosmological model.
       \end{abstract}
	\maketitle
          \tableofcontents

\section{Introduction}\label{sI}
With the advancement of technology, new astrophysical results have posed challenges to well-established theories. However, these advancements have also acted as catalysts for scientific progress and have led to the development of new research that can better explain and respond to these constraints. One of the major discoveries that challenged the traditional Big Bang model was the observation of the accelerated expansion of the Universe \cite {riess1998observational, perlmutter1999measurements}.
To explain this late-time accelerated expansion, two categories of models have emerged: the first class comprises models that incorporate an exotic energy density withing the framework of general relativity commonly referred to as dark energy. The second class consists of modified gravity theories such as  $f(R)$ theory \cite{starobinsky2007disappearing}, $f(T)$ theory \cite{bamba2013effective}, Brans-Dicke theory \cite{sen2001late,errahmani2008late,errahmani2006high}, $f(R,T)$ theory \cite{harko2010thin,bouali2023observational} and Gauss-Bonnet theory \cite{nojiri2005gauss,bouhmadi2011cosmology}.  These theories modify the geomertic part of the Einstein action and offer an alternative explanations for the accelerated expansion without invoking dark energy. By modifying the geometric aspects of the Einstein action, these theories offer a different perspective on the nature of gravity and its role in the expansion of the Universe.
On the other hand, dark energy models, such as the cosmological constant initially proposed by Einstein \cite{einstein1922kosmologische}, also provide a viable explanation for the accelerated cosmic expansion. The cosmological constant, although simple and widely accepted by observations, does require precise fine-tuning to match the observations. To overcome the fine-tuning problem, different dynamic dark energy models have been proposed, including quintessence \cite{martin2008quintessence,  Wetterich:1987fm,ladghami2024thermodynamics}, phantom \cite{dahmani2023constraining,dahmani2023smoothing,bouali2021cosmological,bouali2019cosmological,bouhmadi2015little}, k-essence \cite{sen2001late, Armendariz-Picon:2000nqq}, Chaplygin gas \cite{zhang2006interacting}, holographic dark energy \cite{bouhmadi2018more,sharma2022interacting,george2022holographic,bargach2021dynamical,belkacemi2012holographic,bouhmadi2011cosmology}, and generalized holographic dark energy \cite{granda2009new}.\par
The Generalized Holographic Ricci Dark Energy (GHRDE) model \cite{Xu:2009xi} is a theoretical framework proposed by Xu et al. to explain the accelerated expansion of the Universe. It  extends the holographic dark energy models, which incorporating concepts from both quantum mechanics and gravity.
This model integrates the holographic principle \cite{Hooft, Susskind:1994vu}, which suggests that the maximum entropy in a region of space is bounded by its boundary area rather than its volume. The holographic Ricci dark energy model proposes a fascinating approach where the IR cut-off, the Universe's characteristic length, is considered as the inverse of the Ricci scalar curvature. In this work, we consider an interaction between the energy densities of the cold dark matter (CDM) and GHRDE components to see how the two components of the Universe may mutually influence each other's evolution over the cosmic time. The nature of this interaction can manifest in various forms, each implying a different dynamic between CDM and GHRDE.

In an attempt to compare the GHRDE model with the $\Lambda$CDM model, several cosmological diagnostics and observational tests are commonly employed to assess their viability and consistency with empirical data. In our study, we apply various  diagnostic tools to the GHRDE model, including the  statefinder diagnostic \cite{sahni2003statefinder, Alam:2003sc,mhamdi2023comparing},
$Om$ diagnostic \cite{Shahalam:2015lra, Shafieloo:2012rs}, Statefinder Hierarchy \cite{arabsalmani2011statefinder}, Growth rate of perturbations \cite{mhamdi2024epjc}, analysis by the $\omega_H-\omega_H'$ plane \cite{sahni2008two} and the analysis of the stability by the speed of sound \cite{myung2007instability}.

The statefinder parameters are a set of dimensionless cosmological parameters introduced to characterize the expansion dynamics of the Universe beyond what can be revealed by traditional measures like the Hubble parameter or the deceleration parameter. These parameters are defined in terms of the cosmic scale factor derivatives and their combinations \cite{sahni2003statefinder}, providing additional information about cosmic acceleration. They are denoted by $r$ and $s$ and are derived from the scale factor $a(t)$ and its derivatives with respect to the cosmic time.  The Om-diagnostic, another important diagnostic tool in cosmology, is designed to classify different dark energy models and can robustly distinguish between dynamical dark energy and the cosmological constant.  In order to calculate the $Om$ diagnostic, we need to determine the combination of the Hubble parameters and the cosmological redshift. It is a way to analyze observational data within the context of the $\Lambda$CDM model by plotting the behavior of the dimensionless energy density $\Omega_{\Lambda}$ and $\Omega_{m}$ as a function of redshift. In the aim of providing more detailed and precise diagnostics of the cosmic expansion beyond what the original Statefinder parameters can offer, we extend the concept of the Statefinder parameters $r$ and $s$ by introducing higher-order derivatives of the scale factor $a(t)$. These additional parameters are constructed from the third and higher-order derivatives of the scale factor. In addition to the previous methods, the behavior of GHRDE, is analyzed using $\omega_H-\omega_H'$ plane diagnostic which can provide insights into the nature of dark energy in the GHRDE framework and its potential implications for the cosmic evolution. Finally, we present an analysis of the stability  using the speed of sound within the context of GHRDE which reflects the stability of perturbations in the dark energy fluid. Specifically, it determines how rapidly small density fluctuations in dark energy component propagate.\\

An interaction on the dark sector and within the GHRDE model has been previously analyzed for different goals in \cite{bouhmadi2018more,18,19,20,29,33,34,35}. It is noteworthy that observational evidence has favored interactions within the dark sector \cite{36,37} proving highly valuable in mitigating the coincidence problem. The objective of this study is to conduct a comprehensive analytical examination of GHRDE interacting with Cold Dark Matter (CDM). Furthermore, we aim to compare our model to $\Lambda$CDM by means of the aforementioned diagnostic methods to assess the viability  of the interaction GHRDE with experimental data.\\

The paper is organized as follows: In Sect. \ref{s.II}, we give a short description of the general model set up. In Sect. \ref{s.III}, we present the asymptotic behavior and interaction effects. In Sect. \ref{s.IV}, we introduce different diagnostic tools and analyze our model under consideration according to these tools. In Sect. \ref{conclusion}, we discuss the results and conclusions.

\section{General setup}\label{s.II}

We consider a homogeneous and isotropic Universe described by the Friedmann-Lemaitre-Robertson-Walker (FLRW)  metric

\begin{equation}
    ds^2=-dt^2+a(t)^2\left(\frac{dr^2}{1-kr^2}+r^2d\theta^2+r^2\sin ^2\theta d\phi^2\right),\label{1}
\end{equation}
 where $a(t)$ is the scale factor, $k = 1$, $0$, $-1$ for the closed, flat and open geometries, respectively. The Friedmann equation in a flat FLRW Universe, is expressed as

\begin{equation}
    H^2=\frac{1}{3M_{pl}^2}\rho\label{2},
\end{equation}
 
\noindent where $M_{Pl}$ is the reduced Planck mass. We assume that the total energy density $\rho$, of the cosmic fluid is described by a matter component (including cold dark matter (CDM) and baryonic matter) denoted by $\rho_m$, and a Generalized  Holographic Ricci Dark Energy (GHRDE) component, denoted by $\rho_H$. The expression of the GHRDE density is given by \cite{granda2009new},

\begin{equation}
    \rho_H=3 M_{pl}^2\left (\beta \dot H+\alpha H^2 \right ),\label{3}
\end{equation}

  where $\alpha$ and $\beta$ are dimensionless parameters and the dot denotes derivative with respect to the cosmic time. This equation can be written in the following form

\begin{equation}
   \rho_H=3 M_{pl}^2\left ( \frac{\beta}{2}\frac{dH^2}{dx}+\alpha H^2 \right ), \label{4}
\end{equation}

where $x \equiv \ln(a)$. We have adopted the convention $a_0 = 1$, where the subscript $0$ denotes values at the present time. It is convenient to  the energy densities in dimensionless terms,

\begin{align}
 \Omega_m=\frac{\rho_m}{3M_{pl}^2H^2}, \qquad 
 \Omega_H =\frac{\rho_H}{3M_{pl}^2H^2} = \frac{\beta}{2}\frac{1}{E^2}\frac{dE^2}{dx}+\alpha,
\label{6}
\end{align}

\noindent where $E = H/H_0$ represents the normalized Hubble parameter. The Friedmann equation \eqref{2} can be rewritten as  

\begin{equation}
    1=\Omega_m+\Omega_H,\label{7}
\end{equation}
 and provides the following constraint at the current time,

\begin{equation}
    1=\Omega_m^0+\Omega_H^0.\label{8}
\end{equation}

From the expression of the deceleration parameter, $ q=-[1+(1/E).(dE/dx)]$, and using $E(x=0)=1$, we derive the additional constraint

\begin{align}
 dE/dx|_{x=0}&=\frac{\Omega_H^0-\alpha}{\beta}=-1-q_0,\label{10}
\end{align}

\noindent where $q_{0}$ represents the current value of the deceleration parameter. 




 The observation assures that the Universe is currently in accelerated expansion (i.e. $q_0 < 0$). The holographic parameters are thus constrained by the following inequality

\begin{equation}
    \alpha=\Omega^0_{H}+\beta(1+q_0).\label{13}
\end{equation}

Currently, dark energy and dark matter are the major contributors to the energy sector of the Universe, i.e. $\rho $ can be expresed as,  $\rho = \rho_{m} +\rho_{H}$. We assume that both components of energy density are not conserved separately but interact with each other \cite{interaction,Tsallis2018,statefinder2013}. Therefore, the continuity equations take the form 

\begin{align}
\dot\rho+3H\rho(1+\omega)&=0, \label{141}\\
 \dot\rho_H+3H(1+\omega_H)\rho_H&=-Q, \label{14}\\
  \dot\rho_m+3H\rho_m&=Q, \label{15}
\end{align}

\noindent where the quantity $Q$ reflects the interaction between  CMD and GHRDE energy densities. Moreover, positive value of $Q$ represents  transfer of energy from CDM to GHRDE, and vice versa for negative value of $Q$.

\subsection{General equations}

\noindent From a physical point of view and according to  Eqs. \eqref{14} and \eqref{15}, the interaction is defined by the energy densities involved in the system. In particular $Q$ can  be a function of $\dot{\rho}$ or $\dot{p}$ or a function of the energy density `multiplied by a quantity with the units of inverse time. For convenience, we consider that  $Q = Q(H\rho_m, H\rho_H)$. Given that the value of the interaction parameter $Q$ is small \cite{Q}, a power law expansion of $Q$ in terms of energy densities, at first order terms of this interaction  (linear interaction), $Q$ can be expressed as \cite{george2022holographic}

  

\begin{equation}
    Q\simeq\lambda_mH\rho_m+\lambda_HH\rho_H,\label{20}
\end{equation}
where $\lambda_m$ and $\lambda_H$  are constants.\par
By substituting Eq. \eqref{20} into Eq. \eqref{15} and using Eq. \eqref{6}, one obtains the following equation

\begin{equation}
    \frac{d\Omega_m}{dx}=-\Omega_m\left ( 3+\lambda_H-\lambda_m+\frac{1}{E^2}\frac{dE^2}{dx} \right )+\lambda_H.\label{21}
\end{equation}

On the other hand, the substitution of Eqs. \eqref{6} and \eqref{7} into the Friedmann equation, Eq.\eqref{8}, gives the following differential equation

\begin{equation}
   \frac{dE^2}{dx}=2\left ( \frac{1}{\beta}-\frac{\alpha}{\beta} \right )E^2-\frac{2E^2}{\beta}\Omega_m. \label{22}
\end{equation}
Equations \eqref{21} and \eqref{22} form a system of coupled equations. However, by differentiating the two sides of the equation \eqref{22} with respect to $x$, and using Eq. \eqref{21}, we obtain a second order differential equation in $E^2$

\begin{equation} 
 \frac{d^2E^2}{dx^2}=\frac{(-2 \alpha +\beta  (-\lambda_H+\lambda_m-3)+2)}{\beta}\frac{dE^2}{dx} 
 -2\frac{ (3 \alpha +\alpha  \lambda_H-\alpha  \lambda_m+\lambda_m-3)}{\beta }E^2. \label{23}
\end{equation}
Furthermore, the equation of state (EoS) of the GHRDE energy density, $\omega_H=p_H/ \rho_H$, can be written from Eq. \eqref{14} as

\begin{equation}
    \omega_H=-\frac{1}{\Omega_H}\left(1+\frac{1}{3E^2}\frac{dE^2}{dx} \right ).\label{24}
\end{equation}

\noindent The above equation can be rewritten, using equations \eqref{6} and \eqref{7}, as follows

\begin{align}
\omega_H &=-\frac{1}{3}\left [\frac{3}{\Omega_H}+\frac{2}{\beta}\left(1-\frac{\alpha}{\Omega_H}\right) \right ]. \label{25}
\end{align}
 To investigate the behavior of the Universe in the framework of GHRDE, we examine one of the main solutions of the differential equation \eqref{23} which depends on the parameters $\lambda_H$, $\lambda_m$, $\alpha$ and $\beta$.

\subsection{Expected behavior}

The solutions of the differential equation \eqref{23} depend on its discriminant $\Delta=b^2-4\beta$ $c$,  where 
$b=2\alpha-2+\beta(3+\lambda_H -\lambda_m)$
and 
$c=2(\lambda_m -3+\alpha(3+\lambda_H-)$.\\

To illustrate the aim of this paper, we focus on the specific solution corresponding to the positive discriminant, i.e.
\begin{equation}
   E^2(x)=A\exp (\eta _+x)+B\exp (\eta _-x)\label{26},
\end{equation}
where
\begin{equation}
   \eta _{\pm}= \frac{-b\pm \sqrt{\Delta}}{2\beta}\label{27}
\end{equation}
Using the initial condition $E(x = 0) = 1$, Eq.(\ref{10}) from which $\alpha$ is given in terms of $\beta$, the current values of the dimensionless of GHRDE and  the deceleration parameter, the integration constants $A$ and $B$ can be expressed respectively as follows
\begin{align}
A&=\frac{2(1+q_0)+\eta_-}{\eta_--\eta _+} \label{33}\\
 B&=-\frac{2(1+q_0)+\eta _+}{\eta _--\eta _+} \label{34}.
\end{align}
Moreover, by substituting the solution \eqref{26}  into Eq. \eqref{6}, the dimensionless energy density of GHRDE writes
\begin{equation}
    \Omega_H(x)=\frac{\beta  \left(A \eta_+ e^{\frac{\sqrt{\Delta } x}{\beta }}+B \eta_-\right)}{2 \left(A e^{\frac{\sqrt{\Delta } x}{\beta }}+B\right)}+\Omega ^0_H+\beta  \left(q_0+1\right)\label{36}.
\end{equation}
In the following, we will analyze Eq. \eqref{26} and  discuss its behaviors, which depends on the choices of the holographic parameter $\beta$ and of the interaction constants $\lambda_m$ and $\lambda_H$. 

\section{Observational constraints}\label{s.III}

\begin{figure}[h!]
    \centering
    \subfloat[Strong interaction ($\lambda=0.1$).]{\includegraphics[width=0.5\textwidth]{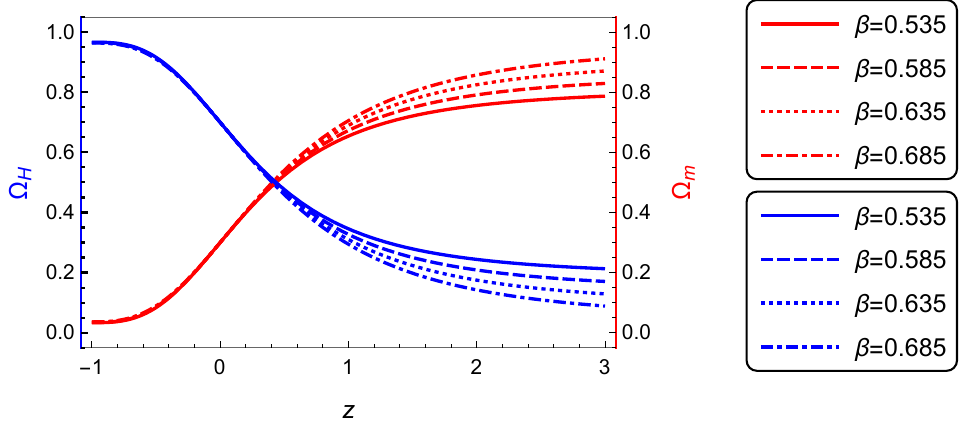}}\label{g.1a}
    \hfill
    \subfloat[Weak interaction ($\lambda=0.0001$).]{\includegraphics[width=0.37\textwidth]{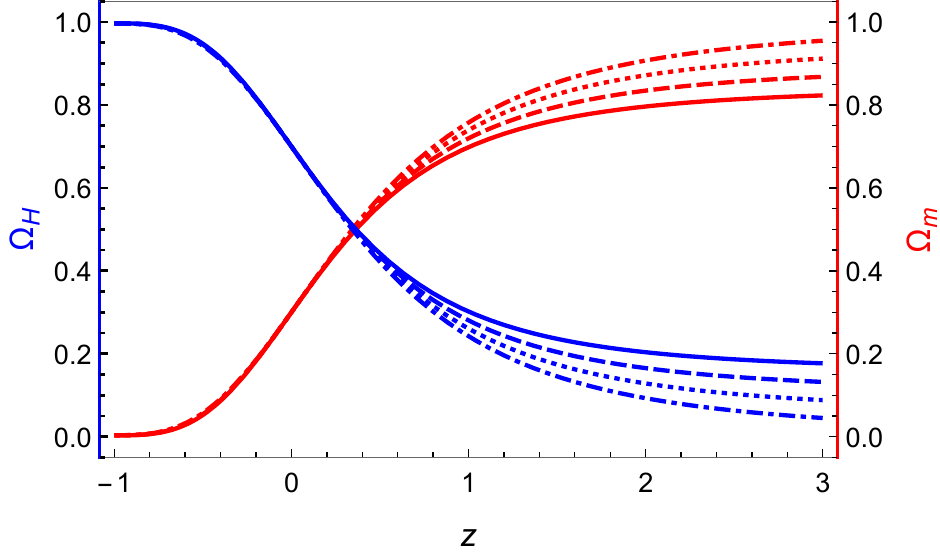}\label{g.1b}}
    \caption{The variation of $\Omega_H$ and $\Omega_m$ as  function of $z$.}
    \label{g1}
\end{figure}

\begin{figure}[h!]
    \centering
    \subfloat[Strong interaction ($\lambda=0.1$).]{\includegraphics[width=0.55\textwidth]{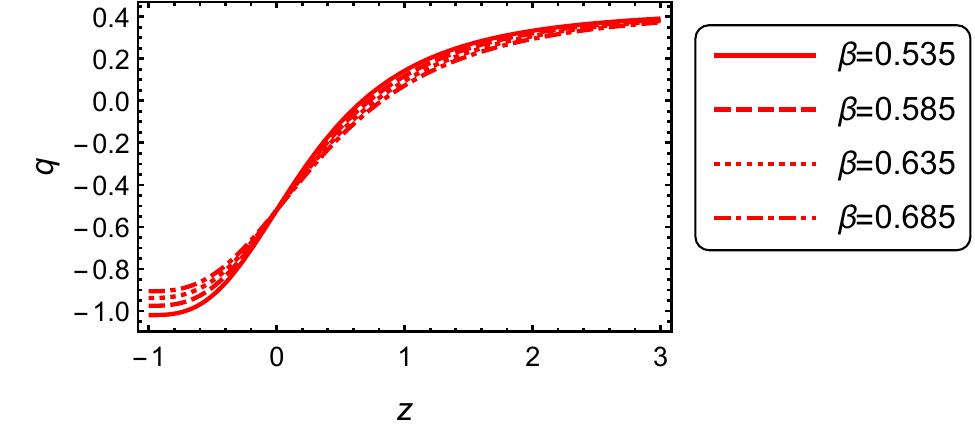}\label{g.1c}}
    \hfill
    \subfloat[Weak interaction ($\lambda=0.0001$).]{\includegraphics[width=0.37\textwidth]{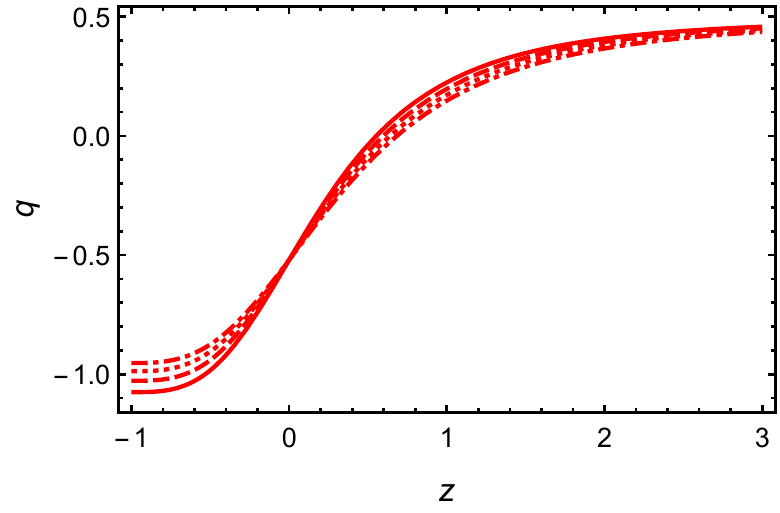}\label{g.1d}}  
    \caption{The variation of deceleration parameter $q$ as a function of $z$.}
    \label{g2}
\end{figure}

\begin{figure}[h!]
    \centering
    \subfloat[Strong interaction ($\lambda=0.1$).]{\includegraphics[width=0.55\textwidth]{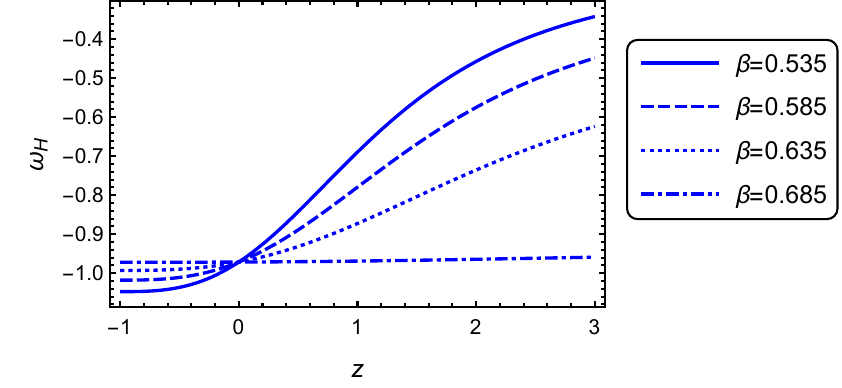} \label{g.1e}}
    \hfill
    \subfloat[Weak interaction ($\lambda=0.0001$).]{\includegraphics[width=0.38\textwidth]{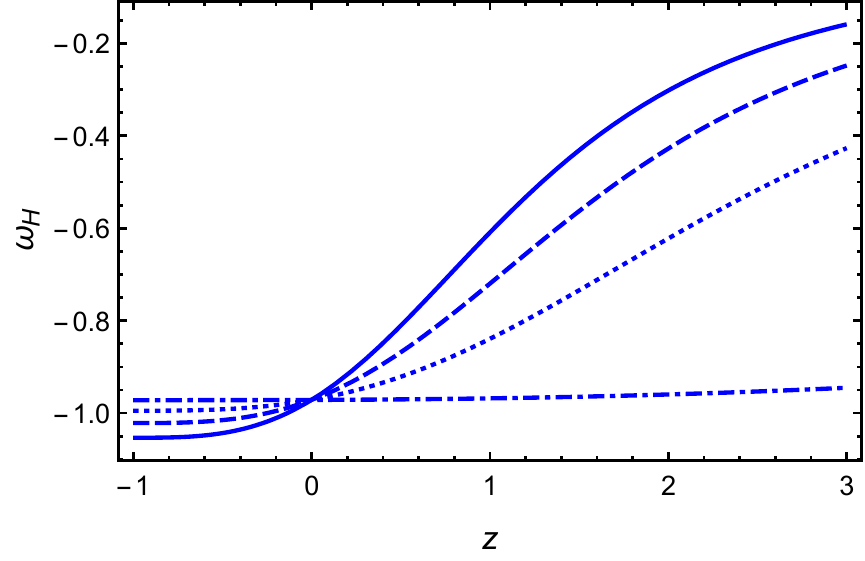}\label{g.1f}} 
    \caption{The variation of $\omega_H$ as a function of $z$.}
    \label{g3}
\end{figure}

In this section, we adopt the current value of dimensionless CDM energy density equal to 0.315 \cite{aghanim2020planck, cunha2008transition}. For the deceleration parameter we adopt the one given by assuming $\Lambda$CDM in \cite{asvesta2022observational} i.e. $q_{0} \simeq -0.527$ . From now on and to illustrate our purpose, we consider that the interaction is such that $\lambda_H=\lambda_m=\lambda$.
In this case, the parameters appearing in Eq. \eqref{27} can be rewritten in terms of the dimensionless dark energy and  the deceleration parameter as follows
\begin{align}
  b&=2\Omega_H^0+5\beta+2(q_0-1) \label{51} \\
    \Delta &=\beta^2(1-2 q_0)^2+4\beta((1-\Omega_H^0)(1-2 q_0)-2\lambda)+4(1-\Omega_H^0)^2\label{52}
\end{align}
To preserve the positivity of the discriminant for all values of $\beta$, we distinguish two classes of solutions. The first class in which the interaction is such that $\lambda > 0.693$ and $\beta$ depends on $\lambda$. The second class is such that $0<\lambda < 0.693$ for all $\beta$. We have plotted the evolution of cosmological parameters, namely the dimensionless dark energy density of GHRDE, the dimensionless  CDM energy density, the equation of state, and the deceleration parameter for both classes. We observe that the first class does not reproduce the area dominated by CDM energy density while the second class does, as shown in Figs. \ref{g1}-\ref{g3}, wi will focus on this second class in the following. \\

Figures. \ref{g1}, \ref{g2}, and \ref{g3}  illustrate cosmological parameters $(\Omega_H,\Omega_m)$, $q$ and $\omega_H$ versus the redshift $z$, respectively. The parameter $\beta$ is chosen to provide the best agreement with observations.  First of all, our model reproduces the accelerated expansion of the Universe in accordance with the recent observations \cite{riess2004type, knop2003new, perlmutter1999measurements}. We also observe that, the transition from decelerated expansion ($q > 0$) to accelerated expansion ($q <0$) occurred for the weak interaction at values of $\beta = 0.535$, $0.585$, $0.635$ and $0.685$ at $z_T =0.553$, $0.597$, $0.640$ and $0. 682$, respectively. These values are in good agreement with the redshift transition of the $\Lambda$CDM model reported in the redshift interval $z_T =\left [ 0.45-0.73 \right ]$  \cite{cunha2008transition}. Concerning the strong interaction, the transition occurred  for the values of $\beta =0.535$, $0.585$, $0.635$ and $0.685$ at $z_T =0.670$, $0.722$, $0.772$ and 0.821, respectively. We clearly see that the first two values are in agreement with those of $\Lambda$CDM, while the two remainning  exceed its upper limit (Figs. \ref{g.1c} and \ref{g.1d}). A  significant remark is that precisely for the weak interaction, the era dominated by CDM is reproduced in the past. Indeed, as shown in Fig. \ref{g.1d}, q tends to 1/2 and hence $\Omega_m=\frac{2}{3}(q+1)$ tends to 1. For the equation of state, Figs. (\ref{g.1e}) and (\ref{g.1f}), show that currently $\omega_H^0\simeq -1$, which is in good agreement with the results of observations \cite{riess2004type}. This indicates that our model reproduces $\Lambda$CDM. Additionnally, it is noted that for both strong and weak interactions, at values of   $\beta$ equal to $0.535$ and $0.585$, $\omega_H$ exceeds the barrier of $-1$. Such an increase could occur in models with two scalar fields in which perturbations remain close to the quintessence case \cite{kunz2007dark}. Alternatively, the phantom divide can be crossed with a sound speed  that vanishes in the phantom domain \cite{creminelli2009effective} or in models with additional terms in the action, such as  kinetic gravity \cite{deffayet2010imperfect}, or with non-minimal coupling \cite{amendola2000coupled, pettorino2008coupled}. Concerning other values of $\beta$ for both interactions i.e. weak and strong, $\omega_H$ remains higher than $-1$ which reflects models with a single minimally coupled scalar field \cite{ma1999mass}. We notice that for models with $\beta =0.685$, $\omega_H$  remains almost constant which means that the pressure and energy density of dark energy mimic a barotropic fluid.\\

\section{Geometrical diagnostics}\label{s.IV}

In this section, we employ various diagnostic tools to compare our model  with tests model.

\subsection{Statefinder analysis}
In this section, we pay attention to the statefinder diagnosis. As we know, the two variables $H$ and $q$  characterizing the expansion of the Universe and  the rate of acceleration/deceleration of the expanding universe respectively are no longer sensitive enough for distinguishing different models since too many cosmological models correspond to the same current value of $q$. To this purpose,  two new parameters are defined by Sahni et al \cite{sahni2003statefinder}, as the statefinder pair $\{r, s\}$.  It can be constructed using both the second and third derivatives of the scale factor $a$. Thus, the  statefinder pair $\{r, s\}$ is defined as,

\begin{align}
r&=\frac{\dddot a}{aH^3}=1+\frac{3}{2E^2}\frac{dE^2}{dx}+\frac{1}{2E^2}\frac{d^2E^2}{dx^2},\label{54}\\
 s&=\frac{r-1}{3\left ( q-\frac{1}{2} \right )}=-\frac{\frac{3}{2E^2}\frac{dE^2}{dx}+\frac{1}{2E^2}\frac{d^2E^2}{dx^2}}{\frac{3}{2E^2}\frac{dE^2}{dx}+\frac{9}{2}}. \label{55}
\end{align}

Using Eqs \eqref{24} and its derivative, these two parameters can be written in terms of the energy density, the equation of state and their derivatives with respect to $x$ as

\begin{equation}
    r=1+\frac{9}{2}\Omega_H\omega_H(1+\Omega_H\omega_H)-\frac{3}{2}(\Omega_H\omega_H)',\label{56}
\end{equation}

\begin{equation}
    s=-\frac{9}{2}(1+\Omega_H\omega_H)+\frac{3}{2}\frac{(\Omega_H\omega_H)'}{\Omega_H\omega_H}.\label{57}
\end{equation}

\noindent The fact that the $\Lambda$CDM model admits a feature given by a fixed point ($r =1$,$s =0$) in the $r-s$ plane, all proposed dark energy models can be tested from a behavioral point of view to see their from that point i.e. from $\Lambda$CDM. To do so, we first plot the statefinder parameters $r$ and $s$ for different values of $\beta$ as a function of redshift in weak and strong interaction.

\begin{figure}[h!]
    \centering
    \subfloat[Strong interaction ($\lambda=0.1$).]{\includegraphics[width=0.58\textwidth]{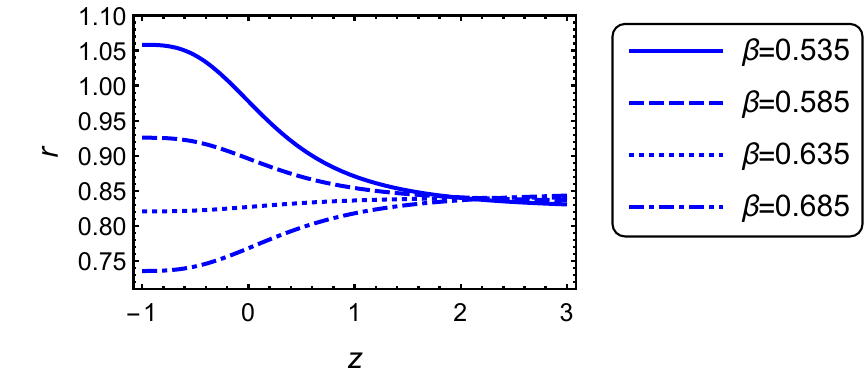} \label{g.4a}}
    \hfill
    \subfloat[Weak interaction ($\lambda=0.0001$).]{\includegraphics[width=0.36\textwidth]{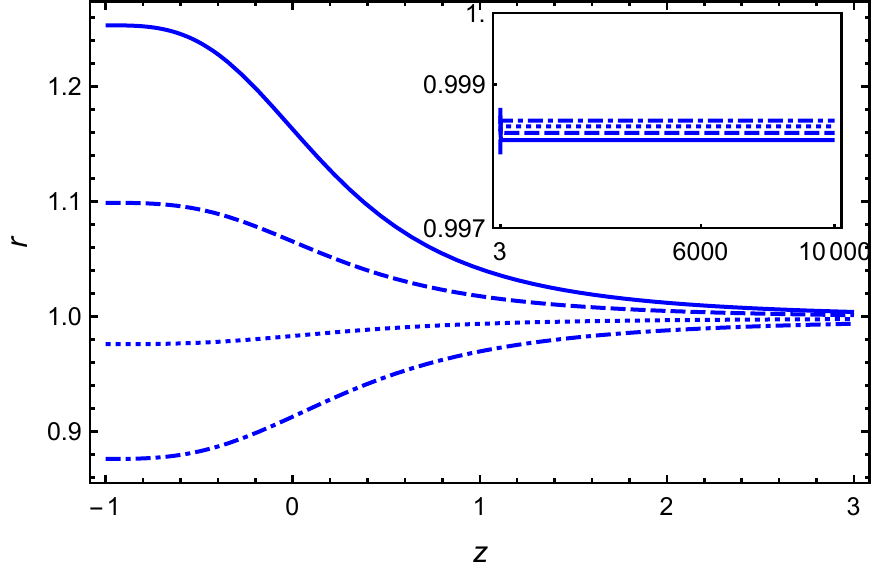} \label{g.4b}} 
    \caption{Evolution of the first statefinder parameter $r$ in terms of redshift.}
    \label{g.4}
\end{figure}

\begin{figure}[h!]
    \centering
    \subfloat[Strong interaction ($\lambda=0.1$).]{\includegraphics[width=0.57\textwidth]{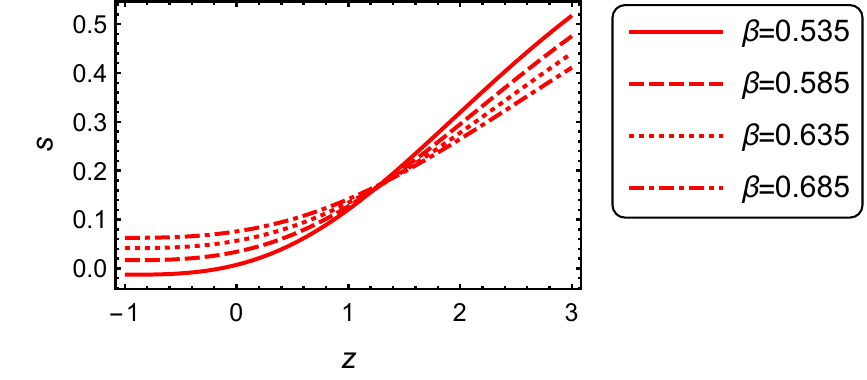}\label{g.5a}}
    \hfill
    \subfloat[Weak interaction ($\lambda=0.0001$).]{\includegraphics[width=0.37\textwidth]{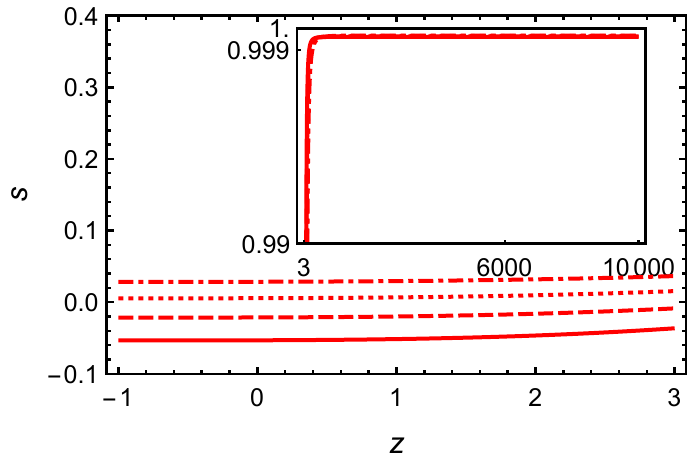}\label{g.5b}} 
    \caption{Evolution of statefinder parameter s} in terms of redshift $z$.
    \label{g.5}
\end{figure}

\begin{figure}[h!]
    \centering
    \subfloat[]{\includegraphics[width=0.4\textwidth]{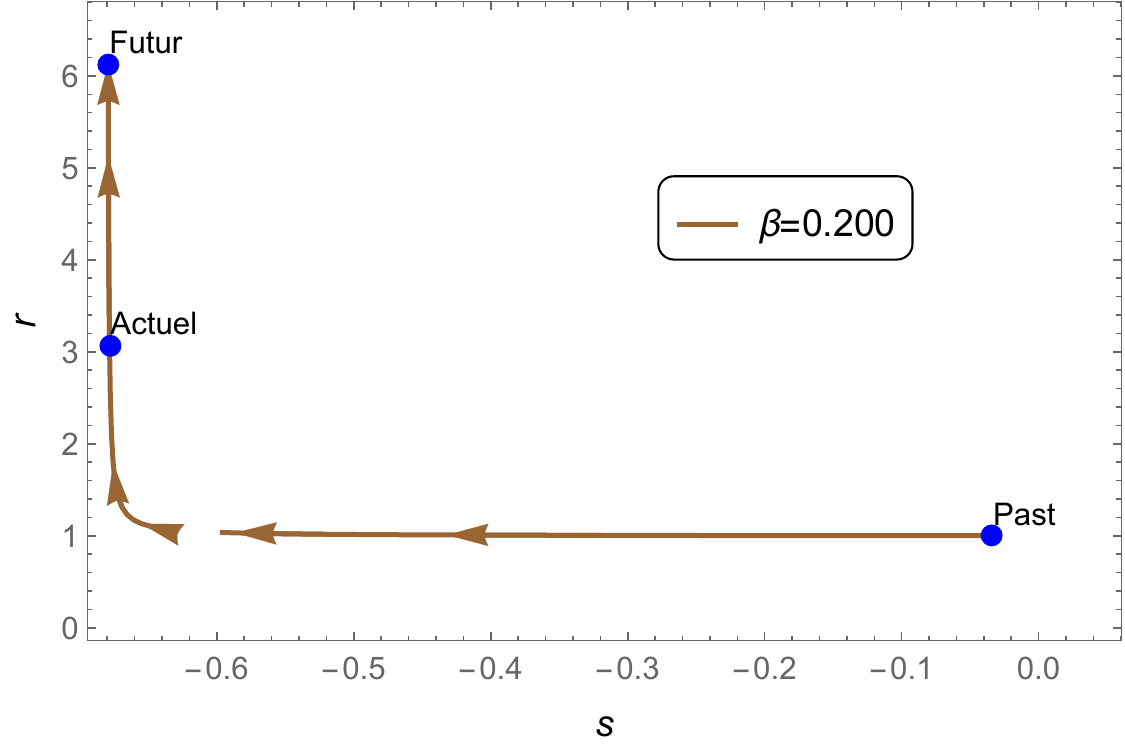}\label{g.6a}}
    \hfill
    \subfloat[]{\includegraphics[width=0.42\textwidth]{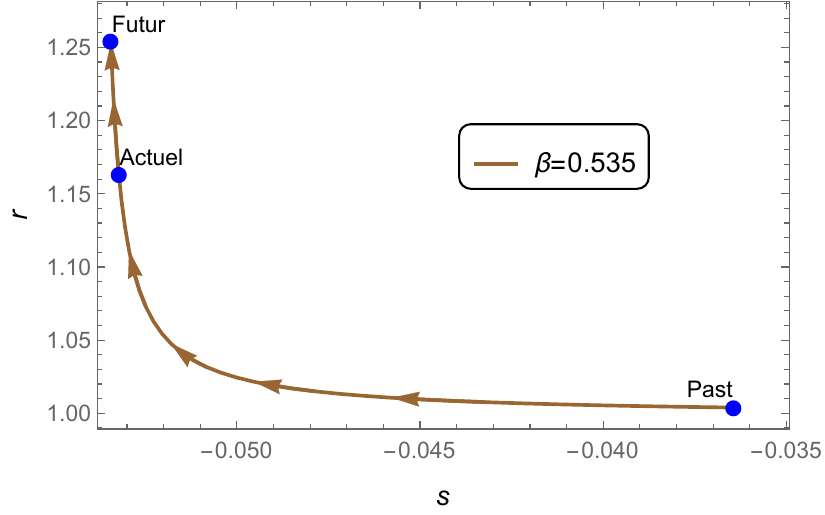}\label{g.6b}}
    \hfill
    \subfloat[]{\includegraphics[width=0.42\textwidth]{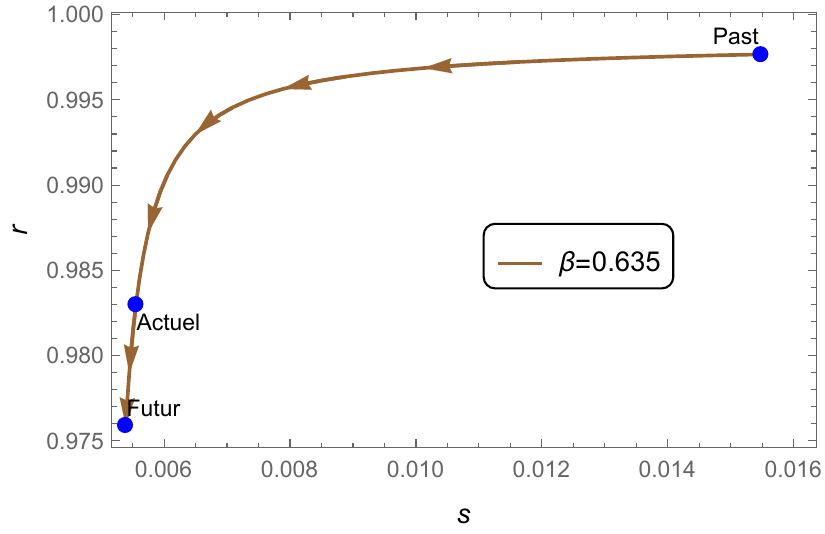}\label{g.6c}}
    \hfill
    \subfloat[]{\includegraphics[width=0.42\textwidth]{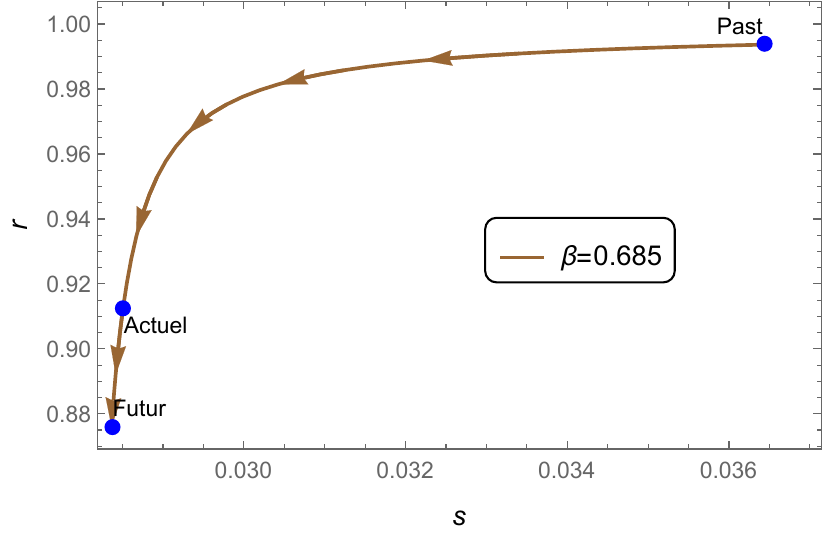}\label{g.6d}} 
    \caption{Evolution of $r$ as a function of $s$ for different values of $\beta$, for the weak interaction $\lambda=0.0001$.}
    \label{g.6}
\end{figure}

\begin{figure}[h!]
    \centering
    \subfloat[]{\includegraphics[width=0.4\textwidth]{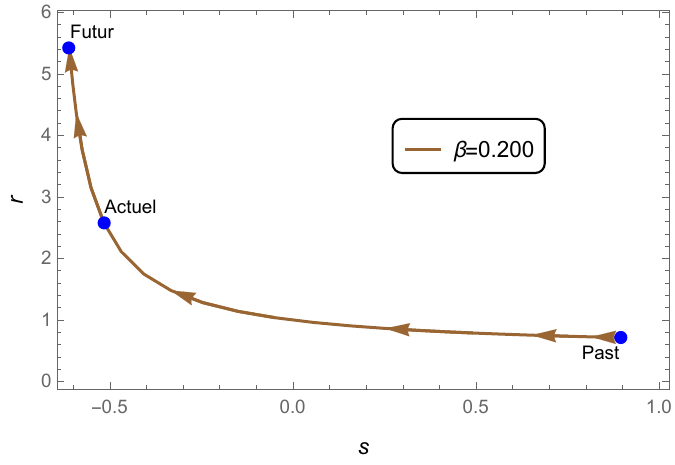}\label{g.7a}}
    \hfill
    \subfloat[]{\includegraphics[width=0.41\textwidth]{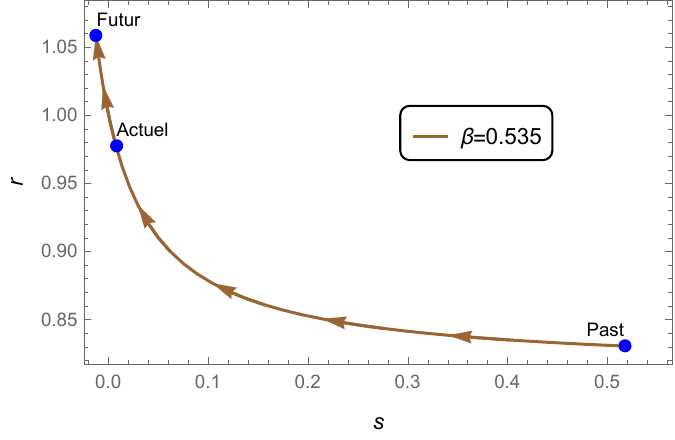}\label{g.7b}}
    \hfill
    \subfloat[]{\includegraphics[width=0.41\textwidth]{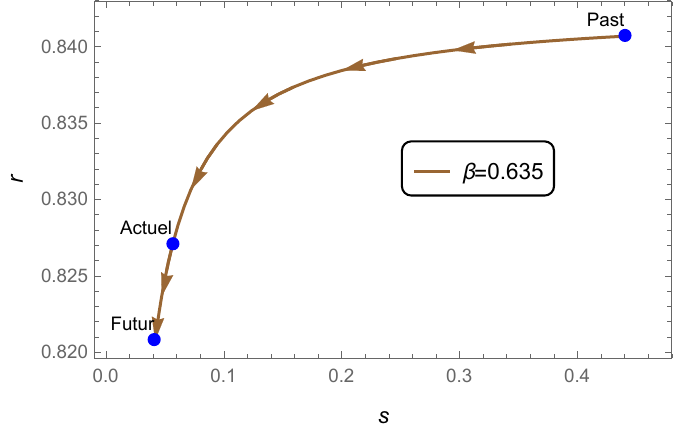}\label{g.7c}}
    \hfill
    \subfloat[]{\includegraphics[width=0.41\textwidth]{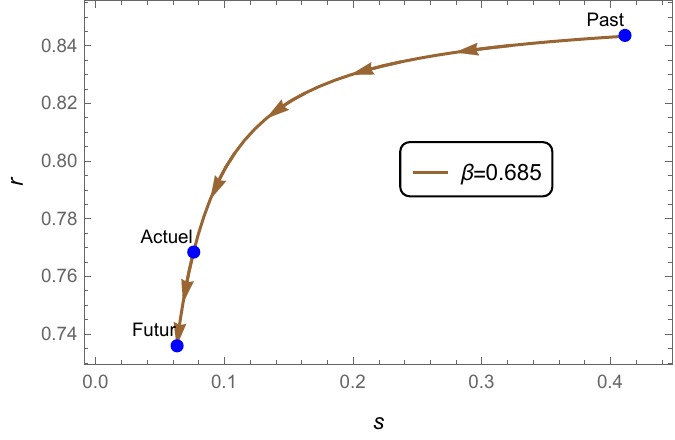}\label{g.7d}} 
    \caption{Evolution of $r$ as a function of $s$ for different values of $\beta$, for the strong interaction $\lambda=0.1$.}
    \label{g.7}
\end{figure}

\noindent Concerning the weak interaction, Fig. \ref{g.6} shows two cases: the case where $r >1$ and $s <0$ which resembles the nature of the Chaplygin gas and the case where $r <1$ and $s>0$ which resembles the nature of the quintessence. The case where  $r\simeq1$ and $s\simeq0$, which mimics the cosmological constant, is also observed in the past. The trajectory of the $\{r, s\}$ point remains very close to the $\{r, s\}$ fixed point in the redshift interval $[-1,3]$, moving towards the past where matter is dominant i.e. our model mimic the standard cold dark matter model (SCDM). For $0 <\beta <0.500$ or $\beta >0.700$, the trajectory of the point $\{r, s\}$ of the model moves away from the fixed point $\{1,0\}$ as shown in Figs (\ref{g.6a}) and (\ref{g.7a}). For the values $\beta=0.200$, $0.535$, $0.635$ and  $0.685$, the current values ($r_0$, $s_0$) are  $(3.076,-0.678)$, $(1.163,-0.053)$, $(0.983,0.005)$ and $(0.913,0.028)$, respectively. From Fig. (\ref{g.7}), the trajectory of the $\{r,s\}$ point in strong interaction moves away from the fixed point of $\Lambda$CDM for all values of $\beta$ except the value $0.535$. The actual values for the pair ($r_0$, $s_0$) for $\beta=0.200,$ $0.535,$ $0.635,$ $0.635$ and  $0.685$ are given by $(2.581,-0.517),$ $(0.978,0.007),$ $(0.827,0.056),$ and  $(0.768,0.076)$, respectively.

\subsection{$Om$ diagnostic}\label{s.V}

$Om(z)$ diagnostic is an effective tool introduced by \cite{sahni2008two} and defined by
\begin{equation}
Om(z)=\frac{E^2(z)-1}{(1+z)^3-1}.
\end{equation}

If the value of $Om(z)$ is  constant at any redshift, the dark energy is exactly in the form of the cosmological constant, while dynamical evolution of $Om(z)$ corresponds to dynamical dark energy models. On the other hand, the slope of $Om(z)$ could distinguish two different types of dark energy models. Positive slope indicates a phantom behaviors ($\omega < -1$) while negative slope represents quintessence ($\omega > -1$) \cite{sahni2008two}.  More recently, a generalized to two-point diagnostic case \cite{huang2011om}
\begin{equation}
Omh^2(z_i,z_j)=\frac{E^2(z_i)-E^2(z_j)}{(z_i+1)^3-(z_j+1)^3}
\end{equation}
 and its improved version of the two-point diagnostic \cite{sahni2014model}  have also been widely used to test $\Lambda$CDM with different $H(z)$ data samples \cite{sahni2014model,ding2015there,zheng2016omh2}.\\

\begin{figure}[h!]
    \centering
    \subfloat[Strong interaction ($\lambda=0.1$).]{\includegraphics[width=0.59\textwidth]{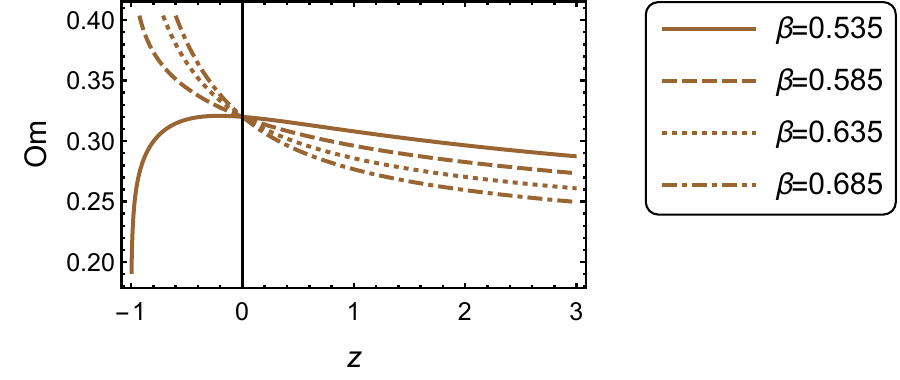}\label{g.8a}}
    \hfill
    \subfloat[Weak interaction ($\lambda=0.0001$).]{\includegraphics[width=0.39\textwidth]{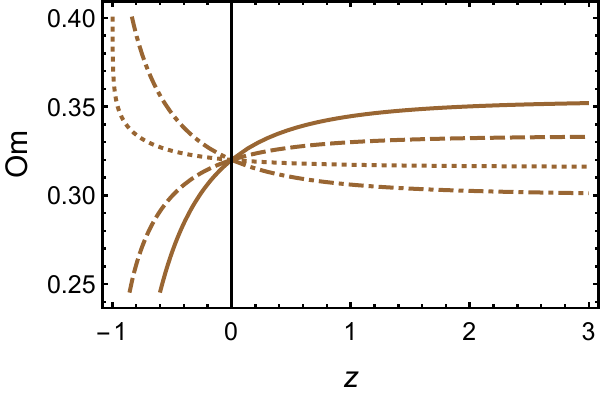}\label{g.8b}} 
    \caption{Evolution of the value of $Om$ in terms of redshift.}
    \label{g.8}
\end{figure}

\begin{table}[htbp]
    \centering
    \begin{minipage}{0.45\textwidth}
        \centering
 \begin{tabular}{|c|c|c|}  
\hline   
($z_i$, $z_j$)&$\beta$&$Omh^2(z_i,z_j)$\\ 
\hline   
\multirow{4}{*}{($0$, $0.57$)}&$0.535$&$0.133\pm 0.002$\\
\cline{2-3}
&$0.585$&$0.132\pm0.002$\\
\cline{2-3}
&$0.635$&$0.130\pm0.002$\\
\cline{2-3}
&$0.685$&$0.128\pm0.002$\\
\hline
\multirow{4}{*}{($0$, $2.34$)}&$0.535$&$0.103\pm0.002$\\
\cline{2-3}
&$0.585$&$0.106\pm0.002$\\
\cline{2-3}
&$0.635$&$0.107\pm0.002$\\
\cline{2-3}
&$0.685$&$0.108\pm0.002$\\
\hline
\multirow{4}{*}{($0.57$, $2.34$)}&$0.535$&$0.100\pm0.002$\\
\cline{2-3}
&$0.585$&$0.104\pm0.002$\\
\cline{2-3}
&$0.635$&$0.106\pm0.002$\\
\cline{2-3}
&$0.685$&$0.106\pm0.002$\\
\hline
\end{tabular}
    \label{t.1a}
    \end{minipage}
    \hfill
    \begin{minipage}{0.45\textwidth}
        \centering
    \begin{tabular}{| c | c | c |}  
\hline   
($z_i$, $z_j$)&$\beta$&$Omh^2(z_i,z_j)$\\ 
\hline   
\multirow{4}{*}{($0$, $0.57$)}&$0.535$&$0.143\pm0.002$\\
\cline{2-3}
&$0.585$&$0.141\pm0.002$\\
\cline{2-3}
&$0.635$&$0.139\pm0.002$\\
\cline{2-3}
&$0.685$&$0.136\pm0.002$\\
\hline
\multirow{4}{*}{($0$, $2.34$)}&$0.535$&$0.120\pm0.002$\\
\cline{2-3}
&$0.585$&$0.122\pm0.002$\\
\cline{2-3}
&$0.635$&$0.124\pm0.002$\\
\cline{2-3}
&$0.685$&$0.124\pm0.002$\\
\hline
\multirow{4}{*}{($0.57$, $2.34$)}&$0.535$&$0.117\pm0.002$\\
\cline{2-3}
&$0.585$&$0.120\pm0.002$\\
\cline{2-3}
&$0.635$&$0.123\pm0.002$\\
\cline{2-3}
&$0.685$&$0.123\pm0.002$\\
\hline
\end{tabular}
    \label{t.1b}
    \end{minipage}
  \caption{Diagnosis developed at two points $Omh^2$ for different values of $\beta$ for strong (left table) and weak (right table) interaction.}
     \label{t.1}
\end{table}

Figure (\ref{g.8}) shows the evolution of the value of $Om$ as a function of redshift. For the strong interaction, Fig (\ref{g.8a}) indicates a negative slope for all values of $\beta$ which mimics a quintessence behavior except for $\beta=0.535$. While for the weak interaction, Fig (\ref{g.8b}) shows negative slope mimicking the quintessence model for 0.635 and 0.685 of $\beta$ values. While the values $0.535$ and $0.585$ mimic phantom models. In Tables (\ref{t.1}), we present the two-point diagnostic using three values of the scaling factor, $H(z=0)=67.4\pm0.5km s^{-1} Mpc^{-1}$ \cite{aghanim2020planck}, $H(z=0. 57)= 92.4\pm 4.5 km s^{-1} Mpc^{-1}$ \cite{sahni2014model} and $H(z=2.34)= 222 \pm 7 km s^{-1} Mpc^{-1}$ \cite{sahni2014model}. The obtained results agree with the one determined by Planck satellite \cite{aghanim2020planck,di2014probing} i.e. $\Omega_{m0}h^2 = 0.1426\pm0.0025$ for the weak interaction and $(z_i, z_j)=(0, 0.57)$.

\subsection{Statefinder hierarchy and growth rate of perturbations}\label{s.V}
In the subsequent discussion and as our focus lies on investigating the late-time evolution of the Universe's expansion, we explore the Taylor expansion of the scale factor around the present epoch $t_0$
\begin{equation}
    a(t)=a(t_0)\left(1+\sum_{n=1}^{\infty}\frac{\alpha_n(t_0)}{n!}(H_0(t-t_0))^n\right),
\end{equation}
 where the first hierarchical statefinder parameter of order $n$, writes
 
 \begin{equation}
    \alpha_n=\frac{1}{aH^{n}}\frac{d^{n}a}{dt^{n}}.
\end{equation}
 
 and the deceleration parameter $q=-\alpha_2$.\\
 
In the $\Lambda$CDM model, $S_n^{(1)}$ is equal to one and it will be used to distinguish between $\Lambda$CDM and other dark energy models. This hierarchical statefinder can be expressed in terms of the first hierarchical statefinder parameter of order $n-1$ by recurrence
\begin{equation}
    S_n^{(1)}=\frac{dS_{n-1}^{(1)}}{dx}+S_{n-1}^{(1)}+\frac{3}{2}S_{n-1}^{(1)}\frac{1}{E^2}\frac{dE^2}{dx}.\label{58}
\end{equation}
where $S^{(1)}_{3}=\alpha_3=r$.
Using Eq. (\ref{56}), we can express the first terms of hierarchical statefinder parameter of order $4$ in terms of $\Omega_H$ and $\omega_H$ as follow
\begin{eqnarray}
 S_4^{(1)}&=&1-9(1+\Omega_H\omega_H)+\frac{99}{4}(1+\Omega_H\omega_H)^2-\frac{81}{4}(1+\Omega_H\omega_H)^3 
+\frac{3}{2}(1+z)\partial_z(\Omega_H\omega_H)\left\{4-\frac{15}{2}(1+\Omega_H\omega_H) \right\}\nonumber\\
&-&(1+z)\partial_z\left\{\frac{3}{2}(1+z)\partial_z(\Omega_H\omega_H)+\frac{9}{4}(1+\Omega_H\omega_H)^2 \right\}.
\end{eqnarray}

The second statefinder corresponding to $S_4^{(1)}$ and defined as \cite{Myrzakulov2013}
\begin{equation}
S_4^{(2)}=\frac{S_4^{(1)}-1}{9(q-\frac{1}{2})}
\end{equation}
can be expressed in terms of $\Omega_H$ and $\omega_H$  as follow 
\begin{eqnarray}
 S_4^{(2)}&=&\frac{1}{\Omega_H\omega_H}\left\{-2(1+\Omega_H\omega_H)+\frac{11}{2}(1+\Omega_H\omega_H)^2-\frac{9}{2}(1+\Omega_H\omega_H)^3\right\}+\frac{1}{3}\frac{1+z}{\Omega_H\omega_H}\partial_z(\Omega_H\omega_H)\left\{4-\frac{15}{2}(1+\Omega_H\omega_H)\right\}\nonumber\\
&-&\frac{1+z}{\Omega_H\omega_H}\partial_z\left\{\frac{1}{3}(1+z)\partial_z(\Omega_H\omega_H)+\frac{1}{2}(1+\Omega_H\omega_H)^2 \right\}.
\end{eqnarray}

The Statefinder parameters can be usefully complemented by the fractional growth parameter $\epsilon(z)$ defined by \cite{acquaviva2010falsify}

\begin{equation}
    \epsilon(z)=\frac{f(z)}{f_{\Lambda CDM}(z)},
\end{equation}
where $f(z) = \frac{d\log \delta}{ d\log a}$ is the growth rate of the structure. Here, $\delta= \frac{\delta \rho_m}{\rho_m}$, with $\delta \rho_m$ and $\rho_m$ being the density perturbation and the energy density of matter, respectively. A good approximation to the growth index is given by \cite{wang1998cluster}
\begin{equation}
    f(z)\approx \Omega_m^{\gamma},
\end{equation}
with
\begin{equation}
     \gamma(z)=\frac{(1-\omega_D)(1-\frac{3\omega_D}{2})(1-\Omega_m(z))}{125(1-\frac{6\omega_D}{5})^3}+\frac{3}{5-\frac{\omega_D}{1-\omega_D}}.
 \end{equation}
In $\Lambda$CDM i.e. $\epsilon(z) = 1$, the growth index is equal to $\gamma = 0.6$ \cite{lahavdynamical}. For other models, $\epsilon(z)$ is different from $\Lambda$CDM, which justify the reason for its use as a diagnostic by applying the composite null diagnostic $\{S_n,\epsilon\}$, where $\{S_n,\epsilon\}=\{1,1\}$ for $\Lambda$CDM \cite{arabsalmani2011statefinder,sharma2020exploring}.\\

\begin{figure}[h!]
    \centering
    \subfloat[Strong interaction ($\lambda=0.1$).\label{g.9a}]{\includegraphics[width=0.5\textwidth]{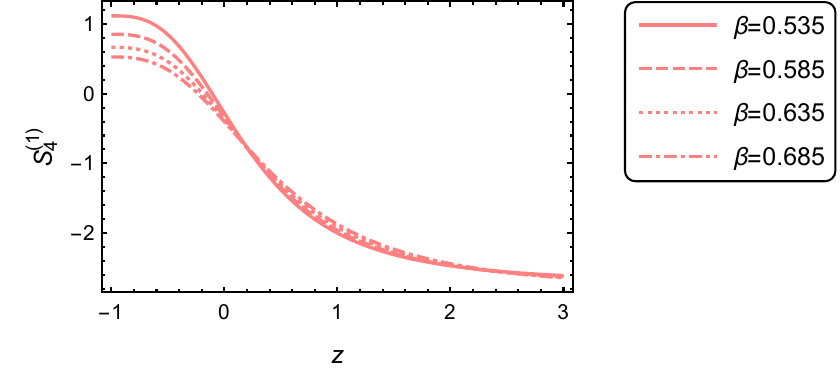}} 
    \hfill
    \subfloat[Weak interaction ($\lambda=0.0001$).\label{g.9b}]{\includegraphics[width=0.35\textwidth]{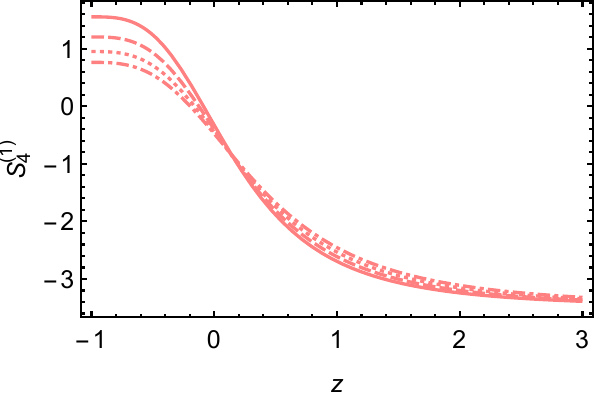}} 
    \caption{Evolution of the first hierarchical statefinder parameter of order $4$, $S_4^{(1)}$, in terms of redshift for strong and weak interaction.}
    \label{g.9}
\end{figure}

\begin{figure}[h!]
    \centering
    \subfloat[Strong interaction ($\lambda=0.1$).\label{g.10a}]{\includegraphics[width=0.5\textwidth]{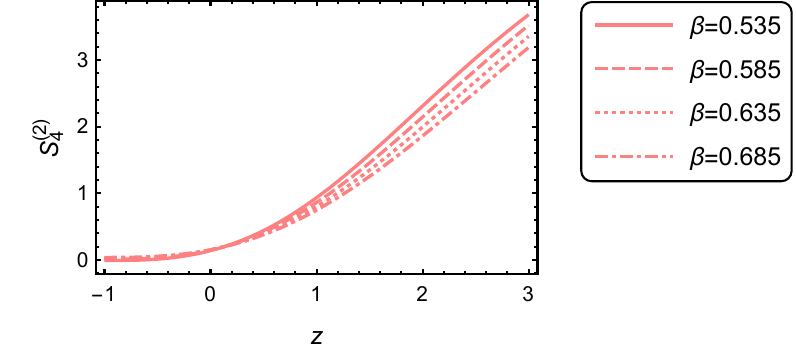}} 
    \hfill
    \subfloat[Weak interaction ($\lambda=0.0001$).\label{g.10b}]{\includegraphics[width=0.35\textwidth]{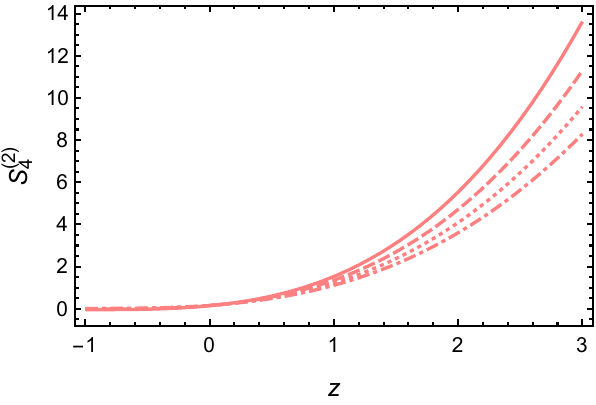}} 
    \caption{Evolution of the second hierarchical statefinder parameter, $S_4^{(2)}$, in terms of redshift $z$ for strong and weak interaction.}
    \label{g.10}
\end{figure}

\begin{figure}[h!]
    \centering
    \subfloat[Strong interaction ($\lambda=0.1$).\label{g.11a}]{\includegraphics[width=0.5\textwidth]{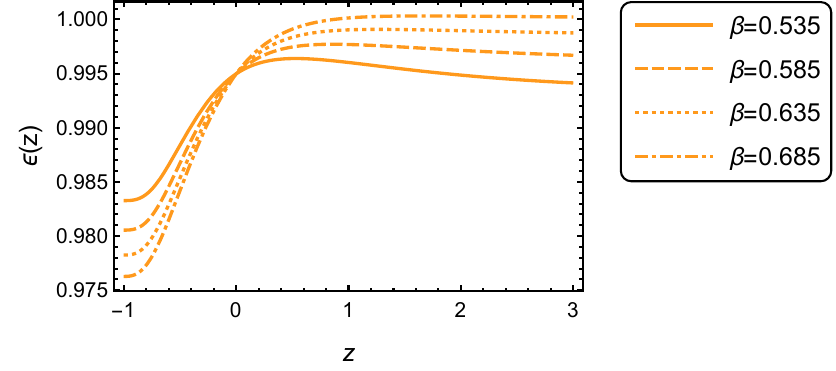}} 
    \hfill
    \subfloat[Weak interaction ($\lambda=0.0001$).\label{g.11b}]{\includegraphics[width=0.35\textwidth]{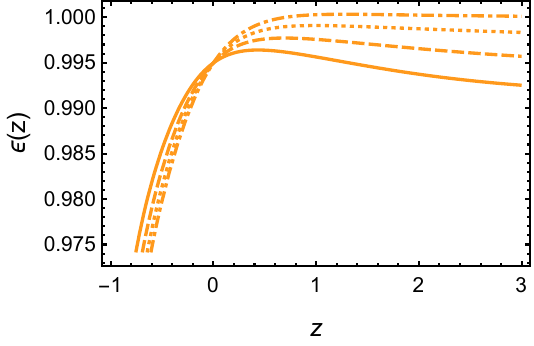}} 
    \caption{Evolution of the growth rate in terms of redshift for strong and weak interaction.}
    \label{g.11}
\end{figure}

\begin{figure}[h!]
    \centering
    \subfloat[Strong interaction ($\lambda=0.1$).\label{g.12a}]{\includegraphics[width=0.4\textwidth]{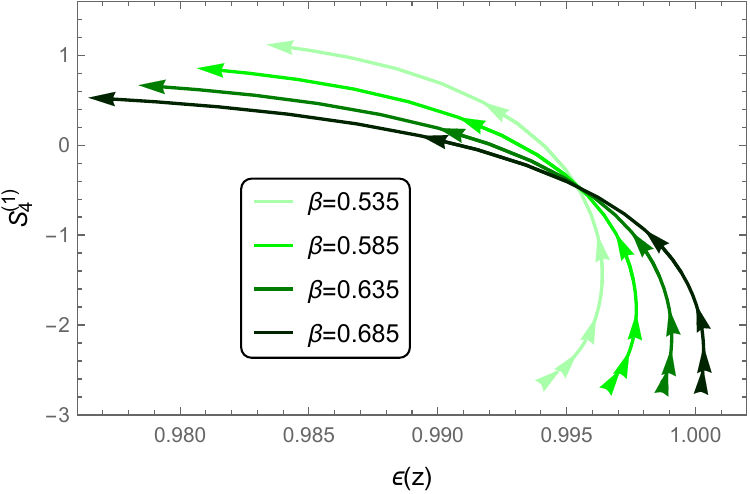}} 
    \hfill
    \subfloat[Weak interaction ($\lambda=0.0001$).\label{g.12b}]{\includegraphics[width=0.4\textwidth]{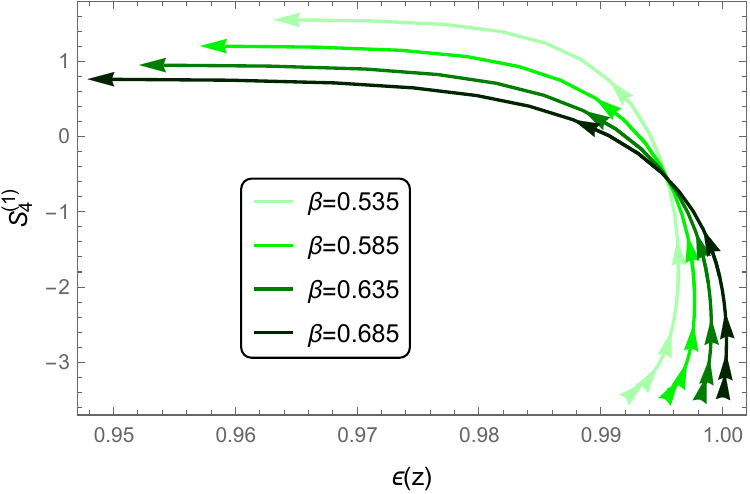}} 
    \caption{Evolution of the first hierarchical statefinder, $S_4^{(1)}$, as a function of the growth rate of the $\epsilon(z)$ structure for different values of $\beta$.}
    \label{g.12}
\end{figure}

\begin{figure}[h!]
    \centering
    \subfloat[Strong interaction ($\lambda=0.1$).\label{g.13a}]{\includegraphics[width=0.4\textwidth]{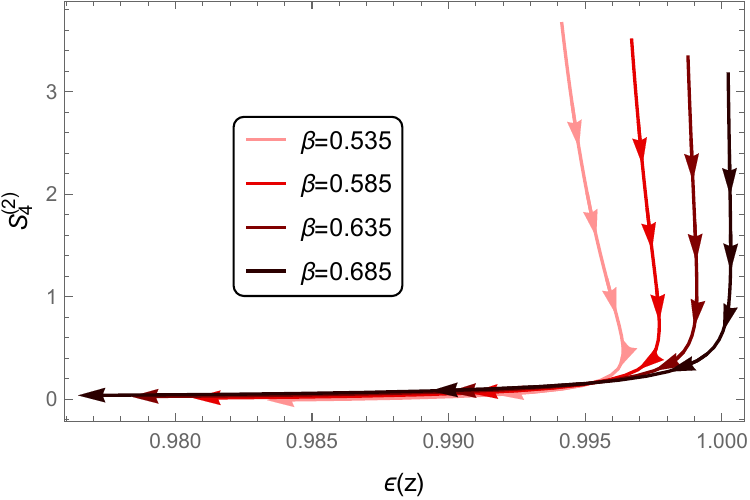}} 
    \hfill
    \subfloat[Weak interaction ($\lambda=0.0001$).\label{g.13b}]{\includegraphics[width=0.4\textwidth]{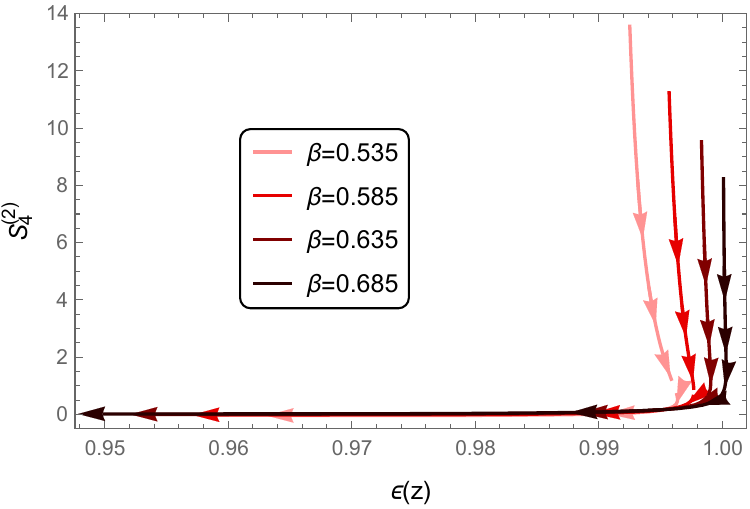}} 
    \caption{Evolution of the second hierarchical statefinder parameter of the fourth order}, $S_4^{(2)}$, as a function of the growth rate of the $\epsilon(z)$ structure for different values of $\beta$.
    \label{g.13}
\end{figure}

Figs. (\ref{g.12}) and  (\ref{g.13}) show that the analysis of the model with the hierarchical statefinder pair of order $4$ and the Growth rate of perturbations in the weak and strong interactions, mimics the cosmological constant model in the future since the pair $\{\epsilon(z),S_4^{(1)}\}$ and $\{\epsilon(z),S_4^{(2)}\}$ reaches approximately $\{1,1\}$ and $\{1,0\}$ respectively. We can clearly see from the figs \ref{g.12b} and \ref{g.13b}, that the value of $\beta=0.635$ gives a composite null diagnostic very close to these two points.

\subsection{Analysis by the $\omega_H - \omega_H'$ plane}

Another useful method that can be added to the various diagnostics is the $\omega_H$ - $\omega_H'$ pair analysis. It was used for the first time by Caldwell and Linder \cite{caldwell2005limits} to classify models into thawing region marked by $\omega_H' >0$ and $\omega_H < 0$, or freezing region marked by $\omega_H' < 0$ and $\omega_H < 0$. Note that the $\Lambda$CDM model is at the $(- 1, 0)$ point. Using Eq \eqref{25}, the variation of the equation of state with respect to $x$  is
\begin{equation}
    \omega_H'=-\frac{\Omega_H'}{\Omega_H} (\omega_H+\frac{2}{3 \beta})
\end{equation}

\begin{figure}[h!]
    \centering
    \subfloat[Strong interaction ($\lambda=0.1$).\label{g.15a}]{\includegraphics[width=0.4\textwidth]{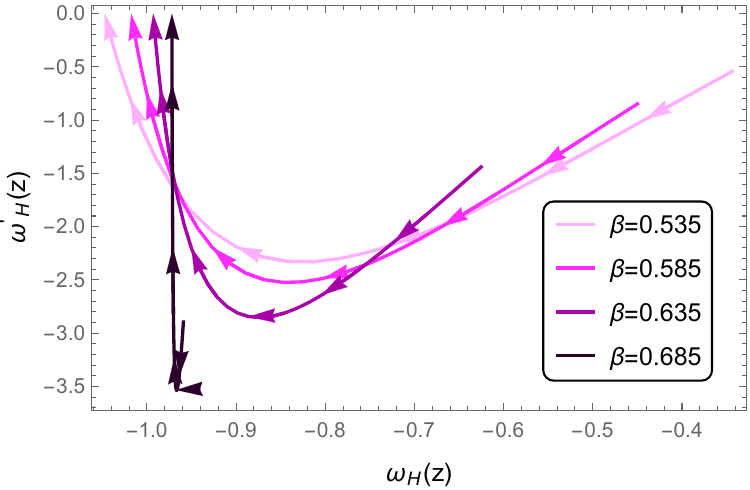}} 
    \hfill
    \subfloat[Weak interaction ($\lambda=0.0001$).\label{g.15b}]{\includegraphics[width=0.4\textwidth]{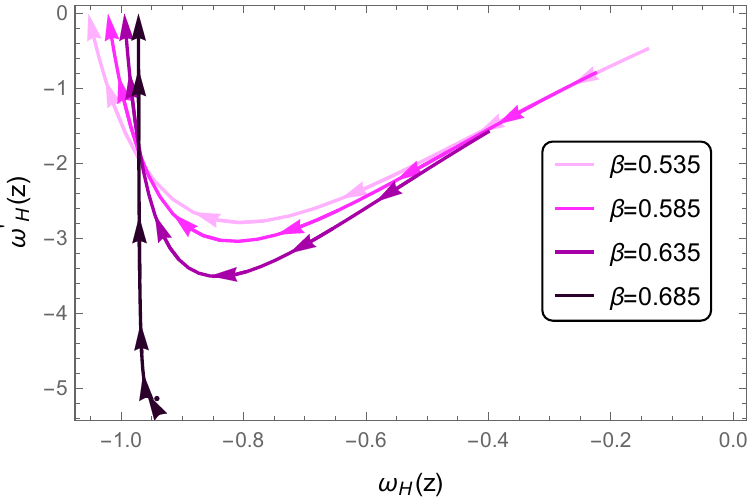}} 
     \caption{Plan $\omega_H^{'}-\omega_H$ for different values of $\beta$ for strong and weak interaction.}
     \label{g.15}
\end{figure}
The $\omega_H^{'}-\omega_H$ plane, defined by the dark energy equation of state parameter and its derivative with respect to the logarithm of the scale factor, is useful for studying the classification of dark energy dynamical models. Fig \ref{g.15} shows that while $\omega_H$ decreases monotonically (Figure \eqref{g3}), $\omega_H^{'}$ retains a negative sign and decreases towards a minimum that varies as a function of the value of $\beta$, before undergoing an increase to reach the value zero.
This represents the region of freezing ($\omega_H^{'}< 0$ with $\omega_H< 0$). The expansion of the Univers is more prounonced than in the region of thawing ($\omega_H^{'}> 0$ with $\omega_H< 0$).

\subsection{Stability analysis}

The speed of sound  is defined by the ratio of the variation, with respect to the cosmic time, of the pressure and dark energy \cite{myung2007instability}. A perfect fluid of dark energy is classically unstable. Indeed an increase of the dark energy means that its pressure becomes more repulsive. Using  $p_H=\omega_H\rho_H$, equations \eqref{6} and \eqref{25}, the speed of sound is given by,

\begin{align}
 v_s^2&= \frac{\dot p_H}{\dot \rho_H}\\
 &=\omega_H+\frac{(1+z)\Omega_H\partial_z\omega_H}{(1+z)\partial_z\Omega_H+3\Omega_H(1+\Omega_H\omega_H)}.
\end{align}

\begin{figure}[h!]
    \centering
    \subfloat[Strong interaction ($\lambda=0.1$).\label{g.16a}]{\includegraphics[width=0.53\textwidth]{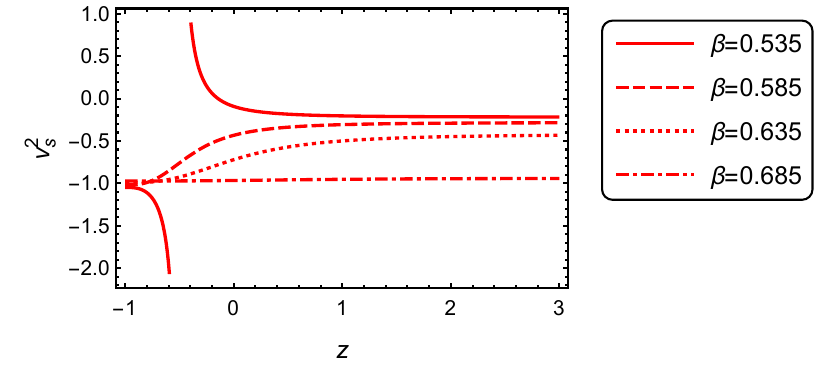}} 
    \hfill
    \subfloat[Weak interaction ($\lambda=0.0001$).\label{g.16b}]{\includegraphics[width=0.34\textwidth]{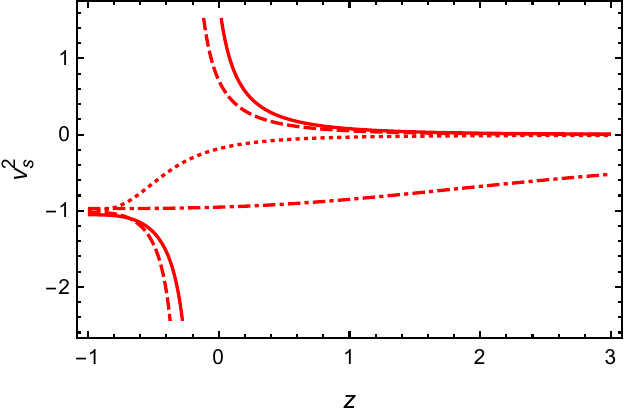}} 
   \caption{Speed of sound $v_s^2$ for strong and weak interaction.}
        \label{g.16}
\end{figure}

From Fig. (\ref{g.16}), we notice that the square of the velocity of the general holographic Ricci dark energy is negative for $\beta=0.585$, $\beta=0.635$ and $\beta=0.685$ for the strong interaction which translates into the instability of the model for these values, while it is stable for $\beta=0.535$. Concerning the weak interaction, the model is stable for $\beta=0.535$, $\beta=0.585$ and $\beta=0.635$ from the past up to now while its unstable for $\beta=0.685$.  

\section{Conclusions}\label{conclusion}
In this paper, we have studied the behavior of the Universe in the framework of GHRDE in the presence of an interaction between the energy densities of the CDM and  GHRDE components. This investigation has allowed for a deeper understanding of the behavior of GHRDE by considering two values of interaction constant $\lambda$ characterizing strong and weak interactions. This can be seen through the analysis of various different cosmological parameters such as the dimensionless dark energy density of GHRDE, the dark matter density, the equation of state, $\omega_H$ and the deceleration parameter, $q$.\\
 
Indeed, for the weak interactions ($\lambda=0.0001$),  we have found that our model reproduces the accelerated expansion of the Universe in agreement with the recent observations. In addition, we havefound that the transition from the decelerated expansion ($q>0$) to the accelerated expansion ($q<0$) took place successively for the values of $\beta= 0.535$, $0.585$, $0.635$ and $0.685$ at $z=  0.553$, $0.597$, $0.640$ and $0.682$, which is in agreement with the redshift of the $\Lambda$CDM model transition in combination with SNIa+CMB data, reported in the interval  $z_T= \left [ 0.45 , 0.73\right ]$. Additionally, we have observed that the Universe corresponds to the era dominated by matter in the past. For the strong interactions ($\lambda=0.1$), the transition from a decelerated to an accelerated phase took place successively for the values of $\beta = 0.535$, $0.585$, $0.635$ and $0.685$ at $z= 0.670$, $0.722$, $0.772$ and $0.821$. This signifies that the two first values are in agreement with that of $\Lambda$CDM, while the two others exceed its upper limit. We also noticed that currently $\omega_{H0} \simeq -1$, which is in agreement with the results of observations. It's noteworthy to hithlight that when $\beta$ equal to $0.535$ and $0.585$, $\omega_H$ exceeds the barrier of $-1$ for both strong and weak interactions. \\

Furthermore, we  have applied various diagnostics tools in order to compare GHRDE with the standard $\Lambda$CDM model, so that one can sheds light on the similarities and differences.  

Initially, we  
 diagnose the model with statefinder method, allowing us to discern three distinct cases:\\- $r> 1$ and $s<0$: this case resembles the nature of Chaplygin gas.\\- $r< 1$ and $s>0$: this case resembles the quintessence.\\- $r \simeq 1$ and $s \simeq 0$: the model mimic the cosmological constant.\\ We noted that the trajectory of the $\{r$, $s\}$ point remains very close to the $\{r$, $s\}$ fixed point in the redshift interval [-1, 3], moving towards the past where matter is dominant. The pair $\{r$, $s\}$ tends to $\{1, 1\}$ which is in agreement with \cite{sahni2003statefinder} as $\Lambda  \simeq 0$.

The second technic employed in this article was the $Om$ diagnostic. The plots showed that for the strong interaction and for all values of $\beta$, the nature of dark energy mimics the quintessence model, except for $\beta=$ 0.535. However, for the weak interaction, the nature of the dark energy mimics the quintessence model for the two values of
$\beta$ =0.635 and $\beta$ =0.685. While for $\beta$ = 0.535 and $\beta$ =0.585, it mimics the phantom model. We mention that the obtained results are in agreement with observational data.\\

Concerning the analysis of our model with Statefinder Hierarchy and Growth rate of perturbations, we have found that for weak and strong interactions the nature of dark energy mimics the cosmological constant model in the future.\\

For the analysis with $\omega_H - \omega'_H$  plane, we noticed a decrease of $\omega_H$ in a monotonic way with a decrease of $\omega'_H$ from a negative value to a minimum which varies according to the value of $\beta$ corresponding to the region of thawing ($\omega'_H>0$ with $\omega_H < 0$).\\

Finally, the analysis of the stability by the speed of sound showed that in the case of weak interaction the model is stable for $\beta=0.535$, $\beta=0.5855$ and $\beta=0.635$ from the past up to now while its unstable for $\beta=0.685$. However, for strong interaction, the model is unstable for $\beta=0.585$, $\beta=0.635$ and $\beta=0.685$ while it is stable for $\beta=0.535$. 


There is no funding in this article.

\section{Data Availability}
No Data associated in the manuscript.

\end{document}